\documentclass[useAMS]{mn2e}
\title{Monopole gravitational waves from relativistic fireballs driving  
gamma-ray bursts}
\author[M. Kutschera]
       {M. Kutschera\\
        Institute of Physics, Jagiellonian University, Reymonta 4, 30-049 
Krak\'ow,  Poland, and, \\Astrophysics Division, H. Niewodnicza\'nski 
Institute of 
Nuclear Physics, Radzikowskiego 152,     31-342 Krak\'ow,  Poland
}
\date{Accepted 2003 July 24.
     Received 2003 July 17;
     in original form 2003 Janurary 12}

\pagerange{\pageref{firstpage}--\pageref{lastpage}}
\pubyear{1994}

\begin{document}

\maketitle

\label{firstpage}

\begin{abstract}

Einstein's General Relativity predicts that pressure, in general stresses,  
play a similar role to energy density, $\epsilon=\rho c^2$ (with $\rho$ being 
corresponding mass 
density), in generating gravity. The source of 
gravitational field, the active gravitational mass density, sometimes referred 
to as Whittaker's mass density, is 
$\rho_{grav}=\rho+3p/c^2$, where $p$ is pressure in the case of an ideal fluid.   
Whittaker's mass is not conserved, hence its 
changes can  propagate as  monopole gravitational waves. Such waves can  be 
generated only by astrophysical sources with varying gravitational mass.    
Here we show that relativistic fireballs, considered in modelling gamma-ray 
burst phenomena,  are likely to radiate monopole gravitational waves from 
high-pressure plasma with varying Whittaker's mass. Also, ejection of 
a significant amount of initial mass-energy of the progenitor contributes to 
the monopole gravitational radiation. We identify monopole waves with
$h^{11}+h^{22}$ waves of Eddington's classification  which
propagate (in the z-direction) together with the energy carried by massless 
fields. We show that the monopole waves satisfy Einstein's equations, with  
a common stress-energy tensor for massless fields. The polarization mode of monopole 
waves is $\Phi_{22}$, i.e. these are perpendicular waves 
which induce changes of the radius of a circle of test particles only (breathing mode).    
The astrophysical importance of monopole gravitational waves is discussed.

\end{abstract}


\section{Introduction}

The coupling of pressure (stresses) to gravity follows from  Einstein's field 
equation,
$R_{\mu \nu}-g_{\mu \nu}R/2=8 \pi G T_{\mu \nu}/c^4$,  where the left-hand 
side (the Ricci tensor, $R_{\mu \nu}$, 
the metric tensor, $g_{\mu \nu}$, and the curvature scalar, $R$) describes the 
geometry of 
space-time, and the right-hand side describes the energy density and stresses 
of the matter distribution in terms of the energy momentum tensor. For an ideal 
fluid, the latter is diagonal in the rest frame of the fluid,  $T_{\mu \nu}=diag(\rho c^2, 
p,p,p)$. Whittaker (1935) has shown that the source of 
gravitational field is $\rho_{grav}=\rho+\Sigma_i T^{ii}/c^2$, where the 
latter term is $3p/c^2$ for an ideal fluid. Under local conditions in the Solar 
system, mass density  
always  dominates, $\rho>> p/c^2$, and the pressure effects  are not noticed.  
The 
gravitational role of pressure is perhaps the greatest in cosmology (Peebles 
1993), where 
the evolution of the scalefactor of the Universe, $a(t)$, is governed by  
Whittaker's mass density,  $3{\ddot a}/a= -4\pi G(\rho + 3p/c^2)$. In 
cosmology, pressure 
and  energy density are comparable, and sometimes it is pressure that 
dominates, as is believed to happen presently in our Universe (albeit with 
negative pressure)(Garnavich et al. 1998). Tolman (1934) in his classic book 
pointed out that the 
gravitational attraction of the photon gas is twice as big as the gravity of 
matter of the same total energy density. This conclusion follows from  
Whittaker's formula for  the photon gas with pressure $p=\rho c^2/3$.

\section{Pressure as a source of gravity of relativistic fireballs}

The high-pressure regime of General Relativity has not been considered 
thoroughly yet (Carlip 1998).  
Here we study the  gravitational role of pressure  for astrophysical objects 
with weak gravitational field. In order for pressure to play any noticeable 
role, the system under consideration should obey a relativistic equation of 
state, i.e. the pressure should be comparable to the energy density, as e.g. for 
the photon gas.  In such a case, all diagonal components of the stress-energy 
tensor are of the same order, $T_{00}\approx3T_{11} \approx 3T_{22} \approx 
3T_{33}$. The most promising 
astrophysical system in this regard seems to be a high-temperature thermal 
pair plasma, $e^+,e^-,\gamma$, invoked in models of gamma-ray bursts 
(Paczy\'nski 1986, Goodman 1986). However, the relativistic equation of state is 
only 
a necessary, but not sufficient, condition for pressure to contribute to  
Whittaker's active gravitational mass density. In massive stars, very hot pair 
plasma comprises the inner core but its presence is not expected to affect the 
gravitational mass. This is because stars are in hydrostatic equilibrium and  
the pressure  gradient is exactly balanced by gravitational forces. The virial 
theorem ensures that the integral of the sum of pressure and gravitational 
stresses vanishes, $\int(3p+\Sigma_i T^{ii}_{grav})dV=0$, as the gravitational 
stress 
contribution is $\int \Sigma_i T^{ii}_{grav}dV=E_{grav} $ and $3\int pdV= 
-E_{grav}$. As the net 
contribution of stresses  vanishes, the active gravitational mass is the total 
energy divided by $c^2$, in spite of very high pressure in stars. As shown by 
Landau \& Lifshitz (1975), the virial theorem is fulfilled also in a more 
general situation, 
namely for any system executing bounded motion.  From conservation of the 
stress-energy tensor, which in the weak-field limit of General Relativity 
reads,  $\partial_{\mu} T^{\mu \nu} = 0 $,  it follows that $ \int \partial_0 
T^{0i}dV=-\int \partial_k T^{ki}dV=-\oint T^{ki} d\sigma_k =0$ as the last 
surface integral vanishes by virtue of the finite extent of the system. Multiplying 
the conservation equation by $x_k$ and integrating over the volume we find, 
after time-averaging, $\int \Sigma_i(T^{ii}_s+ T^{ii}_b )dV=0$, where $T^{ii}_s$ 
and $T^{ii}_b$  are, respectively, diagonal stresses 
due to all particles (including radiation) in the system  and due to  "walls" 
constraining their motion.  

This somewhat abstract discussion shows that gravitational effects of pressure 
could show up only for systems not obeying the virial theorem. Physically, one 
should consider non-stationary systems with unbalanced stresses, such as e.g. 
relativistic pair-plasma fireballs. In recent years, astrophysicists studied 
such fireballs intensively, such as in models of gamma-ray bursts 
(Kobayashi, Piran \& 
Sari 1999).  We consider here 
fireballs of spherical symmetry. Promising models of gamma-ray bursts assume 
as the initial process the quick deposition of a huge amount of electromagnetic 
energy in a small volume in the form of thermal pair plasma. The fireball is 
also loaded with some amount of baryons. High temperature plasma forms a fluid 
as it is opaque to photons as a result of Thomson scattering off electrons and 
positrons. The fireball starts to expand slowly at $t=0$ and then accelerates, 
expanding at a still higher rate. Meanwhile the temperature of the plasma drops and 
accelerated expansion ceases when the temperature is low enough for the $e^+, 
e^-$  pairs to annihilate. According to the estimate given by Goodman (1986), 
plasma becomes 
transparent to photons at $t=t_t$ when the termal energy is $k_B T/m_ec^2 \approx 0.03-0.05$. 
The pressure drops very fast 
and vanishes when the photons can freely escape.

For  fireballs modelling gamma-ray bursts, some baryon load is required.
 The acceleration time, $t_z$, depends on the initial energy per baryon, 
referred to as the random Lorentz factor 
of baryons in the fireball, $\eta$ (Kobayashi et al. 1999), $t_z=\eta 
R_0/c$. At time $t_z$
the fireball enters a coasting phase. For baryon-deficient fireballs, 
$\eta>>1$ and $t_z \approx t_t$. 
 Further 
expansion and interactions with the interstellar medium, while crucial for 
producing the observed gamma rays, will not 
concern us here. The gravitational pressure effects occur  in the early 
acceleration phase.

In the following we discuss for simplicity baryon-free (i.e. pure radiation) 
fireballs, as the presence of baryons does not affect our conclusions, and we 
disregard any surrouding medium, assuming fireballs to be in empty space.
Assuming the initial fireball state to be a uniform plasma 
at rest with initial energy density  $T_{00} = \rho_{in} c^2$, the 
initial 
pressure is $p_{in}=\rho_{in} c^2/3=T_{ii}, i=1,2,3 $. The active 
gravitating mass of the fireball at $t=0$ is 
thus $M_{grav}(0)=V_{in}(\rho_{in}+3p_{in}/c^2)=M_{\gamma}+M_p \approx 
2M_{\gamma}$, where  $M_{\gamma}=V_{in}\rho_{in}$  and $M_p \approx 
V_{in}\rho_{in}$ are, respectively, the mass of  plasma in the fireball 
of initial volume $V_{in}$, and the pressure contribution to the gravitational 
mass.

The gain of a significant 
amount of active gravitational mass during the 
formation  period is a direct consequence of Whittaker's formula.
It is the pressure-generated contribution that grows rapidly and eventually 
levels off. The  
other contribution to the gravitational mass is provided by the total energy 
of the fireball, which, as a 
conserved quantity, remains unchanged. Before the formation of the fireball this 
energy is included in the progenitor mass. Hence the  gravitational mass of 
the fireball,  
composed equally of energy density and pressure contributions, is not a 
conserved quantity. This has  profound consequences  as  it implies emission 
of monopole gravitational waves. 
 We parametrize the changing pressure 
contribution to Whittaker's mass as $M_p =M_{\gamma}F_p(t)$, where the 
function 
$F_p(t)\ge 0$ reflects the time evolution of diagonal components of the 
stress-energy tensor of the fireball.

\section{Monopole gravitational waves resulting from an electromagnetic energy 
burst} 

The effects  of non-conservation of  
pressure-generated gravitational mass, $M_p$, are accompanied by  
gravitational 
effects of the mass-energy ejection of the fireball resulting from vigorous expansion. 
At the end of the acceleration 
period the fireball in the observer's frame forms a thin spherical 
shell of pure radiation expanding with the velocity of light $c$. The thickness of 
the shell is about the initial radius of the fireball, $R_0$ (Kobayashi et al. 
1999). For an observer at 
a distance $r$ from the centre of the fireball, the total energy inside the sphere of 
radius $r$  changes  when the radiation debris leaves the sphere. 
This change of  gravitational mass propagates as  a spherical  
wave. We denote the mass of the progenitor of  the fireball by $M_0$. The 
energy 
of the shell is $E_{shell}=M_{\gamma}c^2$ and this energy flows out of the 
sphere of radius $r$ 
in time $\Delta t=R_0/c$. We can parametrize the change of mass-energy in the 
sphere 
of radius $r$ as a function of time  in the form 
$\Delta M=M_{\gamma}F_m(t-r/c)$, since the shell is a pure electromagnetic 
impulse. The change of mass is $\Delta M=0$ for 
$t-r/c<-\Delta t$, and $\Delta M=-M_{\gamma}$ for $t-r/c>\Delta t$, and thus 
we have 
$F_m(\tau)=0$ for $\tau<-\Delta t$, and $F_m(\tau)=-1$  
for $\tau>\Delta t$. Since the passage time,$\Delta t$, is of the order of 1 s, we 
can approximate 
$F_m(\tau) \approx -\Theta(\tau)$, where $\Theta(\tau)=1$ for $\tau>0$ and 
zero otherwise. Let us note that 
$F_m(\tau) \le 0$.

The gravitational field at a distance $r$ from the centre of the progenitor 
star 
is given by the Schwarzschild metric in the weak-field limit,                  
$ds^2=(-1+2GM_0/rc^2)dt^2+(1+2GM_0/rc^2)dr^2+r^2d\Omega^2$. Birkhoff's 
theorem ensures that, until time 
$t=r/c-\Delta t$, the metric at distance $r$ remains still the same 
Schwarzschild 
metric. When the expanding shell passes  the observer at $r$, at time 
$t=r/c+\Delta t$, the metric changes, as now the mass remaining 
in a sphere of radius $r$ is $M_r=M_0-M_{\gamma}$. 
We can write the metric in the  form    
\begin{equation}
ds^2=(-1+{2GM_0 \over rc^2}+\gamma^{00})dt^2
+(1+{2GM_0 \over \/
rc^2}+\gamma^{ii})dr^2+r^2d\Omega^2, 
\end{equation}
where 
\begin{equation}
\gamma^{00} =\gamma^{ii}=-{2GM_{\gamma} \over c^2} {1 \over r}
\end{equation}
 are corrections due to 
the loss of mass. We are 
forced to conclude that around the time $t=r/c$ a perturbation of the metric 
has 
passed by the observer at $r$ which somewhat "ironed out" the space-time 
($M_r<M_0$). We assume that $2GM_0/rc^2 <<1$ is a small perturbation of the 
Minkowski 
metric in order to obtain the form of the wave analytically, as in the linear 
approximation gravitational fields can be added to one another. However, the 
above argument based on Birkhoff's theorem is valid for a general Schwarzschild 
metric. We can express the propagating metric perturbation in a simple 
form, 
$h_m^{\mu \nu}(r,t)=\delta^{\mu \nu}(2GM_{\gamma}/c^2)F_m(t-r/c)/r$. Here the 
diagonal components $h_m^{\mu \mu}$ are such that, for $t>r/c+\Delta t$, 
$h_m^{\mu \mu}(r,t)=\gamma^{\mu \mu}(r)$ from the metric (1). The 
perturbation $h_m^{\mu \nu}(r,t)$ 
is  a spherical  wave. Using isotropic coordinates and introducing new fields 
${\bar h}_m^{\mu \nu}= h_m^{\mu \nu}-1/2 \eta^{\mu \nu}h_{\alpha}^{\alpha}$, 
we find the diagonal components to be 
\begin{equation}
{\bar h}_m^{\mu \nu}(r,t)=\delta^{\mu 0}\delta^{\nu 0} {4GM_{\gamma} \over 
c^2} {F_m(t-r/c) \over r}.
\end{equation}
The only non-zero diagonal field is ${\bar h}_m^{00}(r,t)$. 
It corresponds to a  monopole wave resulting from sperically-symmetric ejection
of a part of the mass-energy of a gravitating body in the form of an electromagnetic 
burst.

\section{Monopole gravitational waves from changing Whittaker's mass}

The shell of ejected mass-energy will be preceded by the 
monopole wave resulting from  the change of Whittaker's mass in the 
formation phase of the fireball. 
The models of gamma-ray bursts do not specify 
the nature of the engine that energizes the fireball. It could be collapse of 
the core of a massive star to a Kerr black hole, with the initial fireball 
energy 
extracted from the rotational energy of the black hole by magnetic fields (Mac 
Fadyen \& Woosley 1999). 
In any case, the engine is a massive object  with mass, $M_{bh}$, of a few 
solar 
masses, which after formation of the fireball is assumed not to change. For 
actual estimates we use fireball parameters from Kobayashi et al. 
(1999). The initial size of 
the fireball is  $R_0=300000$ km (about one light-second). Suppose that the 
total fireball energy deposited in this volume corresponds to 1/10 of a solar mass, 
$M_{\gamma}=0.1 M_{\odot}$. The mean mass density is $\rho=1.8$ g/cm$^3$. This 
is a rather low mass 
density and certainly the gravity of the fireball is weak. The total gravitational 
mass of the fireball progenitor, including the mass of the engine, is thus 
$M_0=M_{bh}+M_{\gamma}$, which is the active gravitational mass before 
the formation of the fireball. 
At $t=0$, when  the fireball is formed, the initial pressure contribution to 
the 
gravitational mass becomes comparable to the fireball mass, $M_{\gamma}$, for 
baryon-free fireballs. It is 
difficult to assess how closely this value is approached, as it depends on the 
formation process. However, if  the formation process of the pair plasma is 
rapid enough for a sufficiently high density of $e^+, e^-$ charges to be 
produced in 
the whole initial volume, the photons are trapped and initially the bulk of 
the plasma is essentially at rest, except  of the surface layer of thickness 
of the order of the photon mean free path.  The active gravitational mass of such 
a fireball at rest, from Whittaker's formula, is about $2M_{\gamma}$. Thus the 
gravitational mass of the host star and the fireball  grows to 
$M_{grav}(0)=M_0+M_{\gamma} = M_{bh}+2M_{\gamma}$ on a formation time-scale. 
The pressure contribution reaches the 
maximum value at the formation, and 
remains later at the same level, even after the 
fireball ceases to accelerate at $t=t_z$. To see this let us remember that  the 
pressure-generated mass is, for expanding fireball, given by an integral of 
the sum of diagonal stresses, 
\begin{equation}
M_p= {1 \over c^2} \int\Sigma_iT^{ii} 
dV=\int[(\rho+{p \over c^2})\gamma^2 {v^2 \over c^2}+{3p \over c^2}]dV,
\end{equation}
where $v$ and 
$\gamma=1/\sqrt{1-v^2/c^2}$ are, respectively, the radial expansion velocity 
and the Lorentz factor of the expanding 
fluid element. At $t=0, v=0$ and $M_p=M_{\gamma}$. At the end of the acceleration 
phase,
$t=t_z$, pressure vanishes, $p=0$, and $M_p(t_z)=\int\rho \gamma^2v^2/c^2 dV 
\approx \int \rho \gamma^2 dV=\int T^{00}dV/c^2=M_{\gamma}$.

Whittaker's  mass varies as 
$M_{grav}(t)=M_0+F_p(t)M_{\gamma}$, where 
the function  $F_p(t)$ is introduced to describe the evolution of the active 
gravitational mass  of the 
system. By definition,  
$F_p(t)=0$ for $t<0$ as the gravitational mass is then $M_0$. Near $t=0$ the 
function steeply grows to its maximum, $F_p(0)=1$. 
Model 
fireball calculations (Kobayashi et al. 1999, Goodman 1986) set as 
their initial conditions a step-like behaviour 
of the function $F_p(t)$ at $t=0$. In reality, there will be some formation 
time of 
the fireball,  corresponding to steep but continuous growth of $F_p(t)$. This 
changing Whittaker's mass generates gravitational waves, which can be 
calculated from the field equation $\partial_{\sigma}\partial^{\sigma} {\bar 
h}^{\mu \nu}_p=-16\pi G T^{\mu \nu}/c^4$ (Wald 1984). For a spherical monopole 
wave one 
can easily obtain the solution.  
A general form of such a gravitational wave is
\begin{equation}
{\bar h^{\mu \nu}}_p=A^{\mu \nu} {f_p(t-r/c) \over r}.
\end{equation}

The spherical monopole wave (5) that  
satisfies the wave equation in an empty space, $T^{\mu \nu}=0$, far 
from the source, is also a solution of the nonuniform wave 
equation 
\begin{equation}
A^{\mu \nu}\partial_{\sigma}\partial^{\sigma}{f_p(t-r/c)  \over 
r}=T^{\mu \nu}({\bf r},t),
\end{equation}
with the source 
\begin{equation}
T^{\mu \nu}({\bf r},t) = B^{\mu \nu}M_{\gamma}c^2F_p(t)\delta^{(3)}({\bf r}),
\end{equation}
located at $r=0$, where $B^{00}=0$, $B^{\mu \nu}=0$ for $\mu \ne \nu$, and 
$B^{ii}=1/3$. We can 
identify the source function, $F_p(t)$, with the function describing the 
time evolution of the active gravitational mass of the fireball.
Any change of the active gravitational mass propagates with 
the velocity of light, $c$, and hence the function $F_p(t)$ generates  a 
spherical wave $f_p(t-r/c)/r$. In equation (6) the fireball is approximated by 
a point-like source with time-depended gravitational mass. We define it to be 
the mass within a few initial radii of the fireball $R \sim R_0$. The source 
of 
gravitational radiation switches off, $F_p=0$, when the sum of diagonal 
stresses 
vanishes there, $\Sigma_iT^{ii}=0$. This happens when the radiation debris 
leave the sphere of radius $R$. We estimate the duration of the source 
activity to be  $T \sim t_f+t_a+R_0/c$, where 
$t_f$ is the formation time and $t_a$ is the acceleration time for plasma to 
acquire  relativistic velocities (we 
neglect any time-dilation 
effects due to motion of the fireball progenitor). After all debris from the 
fireball leaves the sphere $R$, the 
metric becomes that of  the remnant of the star with  gravitational mass 
$M_{grav}=M_{bh}$, corresponding to $F_m=-1$.  The duration of the gravitational 
impulse is of the order of $\Delta t$,
$T \sim R_0/c$. For $R_0$ used above, $\Delta t=1 s$.

The solution of the wave equation (6) for a point-like source is found to be  
$f_p(t-r/c)=F_p(t-r/c)$.
Since the total 
energy is conserved, there is no time-dependent contribution to the $\mu=\nu=0$ 
component of the perturbation, ${\bar h}_p^{00}=0$. The space components are      
\begin{equation} 
{\bar h}_p^{ij}(r,t)={1 \over 3}{4GM_{\gamma} \over c^2}{\delta^{ij} \over r} 
F_p(t-r/c), i,j=1,2,3,
\end{equation}
and the trace is ${\bar h}_{p \alpha}^{\alpha}= 
(4GM_{\gamma}/c^2)F_p(t-r/c)/r$. The 
perturbation of the Schwarzschild metric due to changing pressure contribution 
to  Whittaker's mass, $h_p^{\mu \nu}(r,t)$, is a 
spherical wave,
\begin{equation}  
h_p^{00}(r,t)={2GM_{\gamma} \over c^2} {F_p(t-r/c) \over r},  
\end{equation}
\begin{equation}
h_p^{ii}(r,t)=-{1 \over 3}{2GM_{\gamma} \over c^2} {F_p(t-r/c) \over r}, 
i=1,2,3. 
\end{equation}
The total gravitational impulse is described by both 
pressure-generated and mass-loss components, and  the 
resulting perturbation of the metric (1) is $\gamma^{\mu \nu}= h_p^{\mu 
\nu}(r,t)+h_m^{\mu \nu}(r,t)$. Let us remark, that the wave (3) 
obtained purely on physical grounds, is a solution of the wave equation 
(6) 
with the source function $T^{\mu \nu}({\bf r},t) = \delta^{\mu 0}\delta^{\nu 
0}M_{\gamma}c^2F_m(t)\delta^{(3)}({\bf r})$.

We wish to emphasize that 
non-conservation of  Whittaker's active gravitational mass does not violate 
any conservation law. Energy and momentum are strictly conserved, as 
the divergence of the stress-energy tensor vanishes,  $\partial_{\mu} T^{\mu \nu}=0$, 
but stresses, such as kinetic energy and 
pressure, are not conserved separately.  In contrast, the ejection of matter, 
which gives rise to $h_m^{\mu \nu}(r,t)$, conserves the 
energy: the gravitational mass within a given radius  changes by the amount 
taken out by fireball 
debris moving out of  the sphere that we consider. The latter example provides a 
physical proof that the wave is not an artefact which can be removed by changing gauge.
One can imagine a gedanken experiment, which is perhaps more suggestive. Let us 
consider small quantity of hydrogen gas of mass $M$, in a container of 
negligibly small mass. 
The metric around this body is everywhere given by a weak-field Schwarzschild 
formula. Let  half of the initial mass be replaced by the same amount of 
antihydrogen, initially separated from hydrogen. The gravity of antimatter is 
thought to be the same as that of matter and the metric still corresponds to 
the same mass $M$. Then at $t=0$ the separating mechanism is switched off, with 
matter and antimatter annihilating each other. A fireball of pure radiation of 
energy $Mc^2$ is formed. Assuming spherical symmetry and efficient 
annihilation, the fireball becomes a thin shell of radiation, similar to that 
discussed in Section 3. Clearly, a monopole gravitational wave generated during 
the formation of the fireball travels
together with the expanding electromagnetic shell that erases the 
Schwarzschild matric. The space-time, after passing the wave, becomes a flat  
one.

\section{Monopole waves propagating with radiation of massless fields:
a unified approach}

The energy conservation condition
\begin{equation}
{\partial \over \partial t} \int T^{00} dV = -{1 \over 2} \int T_{\mu \nu} 
{\partial h^{\mu \nu}_p \over \partial t}dV
\end{equation}
shows that the fireball, during the pressure build-up, radiates energy by 
gravitational waves, 
irrespective of the geometry. This is a major difference in 
comparison with oscillating non-relativistic sources, where spherically-symmetric 
motion does not generate any gravitational radiation. However, to find the nature 
of emitted waves we must
first consider the question of coherence. 

Far from the source, the gravitational impulse can be represented by a 
superposition of plane waves with weights given by the Fourier transform of the 
signal envelope, $F_p+F_m$. As is well known, for waves propagating in the vacuum  
there exist only two independent amplitudes (Wald 1984), corresponding to 
two polarizations of transverse tidal oscillations. 
Let us focus on the pressure
contribution (8) and consider the wave propagating in the z-direction. We can 
write the wave in the transverse-traceless (TT) gauge in the form 
\begin{equation}  
h^{\mu \nu}_{TT}(z,t)={1 \over 2}[{\bar h}^{11}_p(z,t)-{\bar h}^{22}_p(z,t)]{\bf e}^{\mu
\nu}_+,
\end{equation}
where  the matrix ${\bf e}^{\mu \nu}_+$ is the unit tensor of the "plus" polarization. 
Formally, this sum is identically zero, as all the diagonal components ${\bar h}^{ii}_p$ 
in equation (8)
are the same. Physically, however, we can notice that the amplitude (12) vanishes as a result 
of exact cancellation of two waves of precisely opposite polarizations, 
$h^{\mu \nu}_{TT}=A_+{\bf e}^{\mu \nu}_++(-A_+){\bf e}^{\mu \nu}_+$.
This can happen only when the radiation generation is fully coherent. Since then no energy 
is emitted, we conclude that the energy must be radiated away by  
incoherent gravitational radiation.

In the case of violent explosion, we do not expect much coherence, in 
particular at short wavelenghts,
as this would require
suppression of any randomness in the formation of the fireball. 
Thus at high frequencies  $\nu>>{\bar \nu}$, much higher than ${\bar \nu}$, a typical 
frequency ${\bar \nu} \sim 1/\Delta t \sim$1 Hz, we expect the gravitationl radiation
to be incoherent.
The presence of high frequencies (short wavelegths) depends on the
time behaviour of the fireball formation. Very rapid formation could be approximated by an
instantaneous process, with the time dependence of the source energy-momentum tensor (7) 
given by $F_p(t)=\Theta(t)$. The Fourier transform of $T^{\mu \nu}({\bf r},t)$ is 
$S^{\mu \nu}({\bf k}, \omega) \sim i/(2\pi \omega)$. The corresponding energy distribution of 
the emitted radiation in frequency and angle is then (Adler \& Zeks 1975) 
\begin{equation}
{dE \over d\Omega d\omega} \sim \omega^2 S_{ik}^*({\bf k}, \omega)S^{ik}({\bf k}, \omega) \sim
const.
\end{equation}
Clearly, the $F_p(t)=\Theta(t)$ behaviour is unrealistic, as the total energy of gravitational waves
is infinite. It shows, however, that the more rapid the formation the higher frequencies are
involved. Also, total energy radiated away as gravitational waves grows for more rapid formation 
processes (Adler \& Zeks 1975). The physical function $F_p(t)$ provides a natural cut-off frequency. 

When the short-wavelenght gravitational waves are produced abundantly, one should not regard
the emitted gravitational radiation as propagating in a vacuum. In this case, as shown by Isaacson
(1968), we can treat 
high-frequency perturbation in the geometrical optics limit,$h_{\mu\nu}\approx
A_{\mu\nu}\exp(ik_{\alpha}x^{\alpha})$ with suitably defined $A_{\mu\nu}$ and $k$.
The energy of high-frequency gravitational waves should be included as a source term
through an effective "graviton" stress-energy tensor 
\begin{equation}
T^{\mu \nu}_{gw}=\Sigma_{\bf k}q({\bf k})^2 k^{\mu} k^{\nu},
\end{equation}
where $q({\bf k})^2=\epsilon^2 A_{\mu\nu}^*({\bf k})A^{\mu\nu}({\bf k})c^4/32\pi G$. 
The low-frequency 
waves would thus satisfy the Einstein equation 
\begin{equation}
R^{\mu \nu}-{1 \over 2}g^{\mu \nu}R={8\pi G \over c^4}  T^{\mu \nu}_{gw}
\end{equation}
which  is not the empty space equation. Its form is the same as for gravitational waves   
associated with the electromagnetic shell considered in Section 3, when the photon stress-energy
tensor is treated in the geometrical optics approximation, 
$T^{\mu \nu}_{\gamma}=\Sigma_{{\bf k}}q({\bf k})^2 k^{\mu} k^{\nu}$ 
(Lindquist, Schwartz \& Misner 1965).
The same stress-energy tensor can also be used to describe the gravitational waves travelling 
together with
the neutrino burst from a supernova explosion (Misner 1965), for neutrinos assumed to be massless
particles.

Monopole gravitational waves propagating together with the radiation of massless fields are thus 
described 
in a unified way by equation (15), for any type of radiation. For massless fields, the wave 
vectors in 
the stress-energy tensor are null, $k_{\mu}k^{\mu}=0$, and the curvature scalar in (15) is $R=0$.
When the radiation propagates in the z-direction, the only components of the stress-energy tensor are
$T_{00}, T_{0z}$, and $T_{zz}$. Classification of polarization modes of gravitational waves, 
given by Eardley, Lee \& Lightman (1973) shows that  in this case the mode 
$\Phi_{22}=-R_{x0x0}-R_{y0y0}$ is non-zero,
where $R_{i0j0}$ are the so-called electric components of the Riemann tensor.
This is a monopole breathing mode
which can be identified with the $h^{11}+h^{22}$ wave of  Eddington's classification 
(Eddington 1960). The monopole polarization differs from "plus" and 
"cross" polarizations of vacuum gravitational waves, and corresponds to a circle of test particles 
changing its radius and preserving circular shape in the plane perpendicular to the 
propagation direction of the wave.

\section{Discussion}

The monopole radiation arising from
the time-dependent pressure contribution to the gravitational mass 
would probe the general relativity sector not yet tested empirically, in 
which foundations of cosmology are rooted. 
This radiation would also encode  
valuable astrophysical information, transmitted directly from inside  the 
relativistic 
fireballs formed in gamma-ray burst sources.

It is expected that  in other 
phenomena involving relativistic fireballs, monopole gravitational waves are 
also emitted. In supernova explosions, high-pressure neutrino fireballs are formed, 
which would emit gravitational waves in a very similar manner to the plasma 
fireballs discussed here. One should  keep in mind that, in astrophysical phenomena where
gravitational waves are thought to be produced, usually a lot of energy is radiated
away by massless fields, photons, neutrinos, and high-frequency gravitons. This could make 
generation of pure vacuum gravitational waves 
with only "plus" and "cross" polarizations less frequent than expected and the real
gravitational signal could involve a significant monopole $\Phi_{22}$ contribution. 

The monopole polarization that we have discussed is the same as predicted in scalar-tensor theories. 
The structure of the Ricci tensor in equation (15), for gravitational waves in the radiation background,
is the same as for vacuum gravitational waves in Brans-Dicke theory (Brans \& Dicke 1961). The 
stress-energy tensor for
photons, neutrinos, and Isaacson's effective stress-energy tensor for gravitons play the same role
in equation (15) as the scalar field term in the Brans-Dicke theory for gravitational waves 
propagating in a vacuum. This fact would
make testing Brans-Dicke theory more difficult.

When some major inhomogeneity is involved in the formation of the fireball, resulting in
the anisotropy of pressure, then the diagonal stresses, $B^{ii}$, equation (6), can differ from one 
another, say
$B^{11}_p>B^{22}_p$. A coherent TT-wave can then be emitted, 
\begin{equation}  
h^{11}_{TT}=-h^{22}_{TT}= {1 \over 2} ({\bar h}^{11}_p-{\bar h}^{22}_p ).   
\end{equation}
This amplitude is formally the same as that for a time-dependend mass quadrupole. 
The gravitational wave detector 
response to such a
gravitational wave would be similar to that for quadrupole waves of the same 
amplitude and frequency.

\section*{Acknowledgments}

This research was partially supported by the Polish State Committee for 
Scientific Research under grants nos. PBZ-KBN-054/P03/02 and 2P03B 110 24,
and through a research grant from the Institute of Physics of Jagiellonian University.
The author is grateful to the anonymous referee for pointing out the problem of 
polarization of monopole waves.


\begin{thebibliography}{99}
\bibitem{b13} Adler R. J., Zeks B., 1975, Phys. Rev. D, 12, 3007
\bibitem{b18} Brans C., Dicke R. H., 1961, Phys. Rev., 124, 925 
\bibitem{b1} Carlip S., 1998, Am. J. Phys., 65, 409
\bibitem{b17} Eardley D. M., Lee D. L., Lightman A. P., 1973, Phys. Rev. D, 8, 3308
\bibitem{b19} Eddington A. S., 1960, The Mathematical Theory of Relativity. University Press,
Cambridge, UK 
\bibitem{b3} Garnavich P. M. et al., 1998, ApJ, 509, 74
\bibitem{b4}  Goodman J., 1986, ApJ, 308, L47
\bibitem{b14} Isaacson R. A., 1968, Phys. Rev., 166, 1272
\bibitem{b5} Kobayashi S., Piran T., Sari R., 1999,  ApJ, 513, 669
\bibitem{b6} Landau L. D., Lifshitz E.M.,1975,  The Classical Theory of 
Fields. Pergamon, Oxford, UK
\bibitem{b15} Lindquist R. W., Schwartz R. A., Misner C. W., 1965, Phys. Rev., 137, B1364 
\bibitem{b7}  Mac Fadyen A. I.,Woosley S. E., 1999, ApJ, 524, 262
\bibitem{b16} Misner C. W., 1965, Phys. Rev., 137, B1360
\bibitem{b8} Paczy\'nski B., 1986, ApJ, 308, L43
\bibitem{b9} Peebles P. J. E., 1993, Principles of Physical Cosmology. 
Princeton Univ. Press, Princeton, NJ
\bibitem{b10}Tolman R. C., 1934, Relativity, Thermodynamics and Cosmology. 
Clarendon, Oxford, UK
\bibitem{b11}  Wald R. M., 1984, General Relativity. Univ. of Chicago Press, 
Chicago, IL
\bibitem{b12} Whittaker E. T., 1935, Proc. Roy. Soc., A 149, 384 
\end{thebibliography}
\end{document}